\documentclass[prb,amsmath,amssymb,floatfix,twocolumn]{revtex4-1} 

\usepackage{natbib,hyperref}

\def\ocite#1 {{\onlinecite{#1}} }

\usepackage[latin1]{inputenc}
\usepackage{amsmath}
\usepackage{amsfonts}
\usepackage{amssymb}
\usepackage{graphicx}
\usepackage{setspace}
\usepackage{color}
\usepackage{epstopdf}
\usepackage{subfigure}

\newcommand{\be}{\begin{equation}}
\newcommand{\ee}{\end{equation}}

\def \ra{{\rightarrow}}

\def \be{\begin{equation}}
\def \ee{\end{equation}}
\def \ba{\begin{array}}
\def \ea{\end{array}}
\def \bea{\begin{eqnarray}}
\def \eea{\end{eqnarray}}
\def \nn{\nonumber}

\def \e{{\epsilon}}

\def \a{{\alpha}}
\def \t{{\theta}}
\def \b{{\beta}}

\def \D{{\Delta}}
\def \d{{\delta}}
\def \w{{\omega}}
\def \s{{\sigma}}

\def \e{{\epsilon}}

\def \nd{{^{\vphantom{\dagger}}}}
\def \yd{^\dagger}
\def \av#1{{\langle#1\rangle}}
\def \ket#1{{\,|\,#1\,\rangle\,}}
\def \bra#1{{\,\langle\,#1\,|\,}}
\def \braket#1#2{{\,\langle\,#1\,|\,#2\,\rangle\,}}

\def \rar{{\rightarrow}}

\def \ra{{\rangle}}
\def \la{{\langle}}

\def \beas{\begin{eqnarray*}}
\def \eeas{\end{eqnarray*}}

\def \GS{{\ket{\Psi_{\rm gs}}}}

\newcounter{indice}

\def \ct{{\cos\theta}}
\def \st{{\sin\theta}}

\def\ve{{\varepsilon}}

\begin{document}

\title{Dynamical probing of a topological phase of bosons in one dimension}
\author{Emanuele G. Dalla Torre}
\affiliation{Department of Physics, Harvard University, Cambridge, MA 02138, USA}

\begin{abstract}
We study the linear response to time-dependent probes of a symmetry-protected topological phase of bosons in one-dimension, the Haldane insulator (HI). This phase is separated from the ordinary Mott insulator (MI) and density-wave (DW) phases by continuous transitions, whose field theoretical description is here reviewed. Using this technique, we compute the absorption spectrum to two types of periodic perturbations and relate the findings to the nature of the critical excitations at the transition between the different phases. The HI-MI phase transition is topological and the critical excitations possess trivial quantum numbers: they correspond to particles and holes at zero momentum. Our findings are corroborated by a non-local mean-field approach, which allows us to directly relate the predicted spectrum  to the known  microscopic theory.
\end{abstract}

\maketitle

\section{Introduction}

According to the Landau paradigm, any two phases with the same symmetries can be adiabatically connected through some (complicated) path in parameter space (the best known example being the liquid and gas phases of water). For many years, the only counter-example to this paradigm was the quantum Hall states, which do not break any local symmetry, but nevertheless cannot be connected together without crossing a phase transition. In recent years, this phenomenon has been extended to a broader class of phases, known as ``topological band insulators'' \cite{KaneMele,Zhang,Qi,Andreas}. Lacking a local order parameter, topological insulators may possibly be used to realize quantum circuits unaffected by local sources of dissipation, such as disorder and time-dependent noise \cite{Zhang2}. However, for the same reason, their experimental characterization is generally very hard: local probes are unable to distinguish between the bulks of a conventional insulator and of a topological insulator.

This problem is usually overcome by studying interfaces between topological phases and trivial phases (such as the vacuum), where protected edge states must appear. In the case of the integer quantum Hall states, the number of edge modes coincides with the number of filled Landau levels and leads to a quantized transverse conductance. Similarly, the appearance of topologically-protected zero-energy modes at the boundaries between different phases constituted the first experimental evidence of the topological insulators \cite{Hasan,Shen,Hasan-rev}. A second strategy consists in measuring non-local ``string'' order parameters, characteristic for example of topological phases in one dimension. A proof-of-principle experiment has already been successfully demonstrated using ultracold atoms in optical lattices \cite{Bloch-string}, following our predictions \cite{PRL,PRB}. 

In the context of ultracold atoms, a third alternative method consists in probing the dynamic response of the system in the vicinity of a transition between different phases \cite{Stoferle,inguscio09}. Here, we specifically refer to continuous phase transitions, which can be detected by the closing of a gap in the response to a local probe \footnote{Local probes do not couple distinct ground states across a first-order transition. As a consequence, the closing of a spectral gap in the response to a local probe is always associated with a continuous phase transitions.}. In the vicinity of these transitions, the low-frequency excitations hold information about the nature of the transition. In particular, because transitions between topologically trivial and non-trivial phases do not break any local symmetry, their low-frequency excitations must possess trivial quantum numbers. In contrast, if a transition breaks a particular symmetry, its low-frequency excitations must possess the same quantum number as the order parameter of the symmetry broken phase. Dynamical probes couple to these low-frequency excitations and may provide useful information about the nature of the transition.

In the case of translational invariance, for example, at the transition from a homogeneous phase to a density-wave phase, the critical excitation must possess the same momentum as the density-wave order parameter. In contrast, at a topological phase transition the critical excitations always possess zero momentum. Similarly, systems that conserve the total number of particles possess an additional $U(1)$ ``gauge'' symmetry, which is spontaneously broken at the transition to a superfluid phase. At this transition the critical excitations are Bogoliubov quasi-particles, which do not conserve the total number of particles (or to be more precise, shift the number of particles by a non-integer amount). On the contrary, at a transition between a normal and topological phase, the symmetry is preserved and, accordingly, the critical excitations must carry integer charge. 

In this article we consider a particular example of one dimensional symmetry-protected topological insulator, the ``Haldane Insulator''  \cite{PRL,PRB,Pollmann,Pollmann2}. This phase arises in models of interacting bosons on a one-dimensional lattice, which are relevant to ultracold dipolar atoms  \cite{Pfau} or molecules  \cite{Jin} and Rydberg atoms  \cite{Buchler}, in optical lattices  \cite{Bloch-rev}. 
These systems support additional phases that are topologically trivial, such as the Mott insulator (MI) and the density-wave (DW) phase. In the simplest DW phase the density oscillates between even and odd sites, spontaneously breaking the discrete translational invariance, modulo two lattice sites. As we will see, the dynamic response demonstrates that the critical excitations of the HI-DW transition are phonons at momentum $k=\pi$. In contrast, at the topological MI-HI transition the dynamic response demonstrates that the critical excitations are simply particles and holes at zero momentum. 

This article is organized as follows. In Sec. \ref{sec:back} we describe the ground state properties of the Haldane insulator and present new numerical results aimed to characterize its low-lying excitations. In Sec. \ref{sec:methods} and \ref{sec:results} we introduce the field theoretical description of the Haldane insulator and use this method to predict the dynamic response of the system. Our findings are in contrast with the predictions of Ref.~\onlinecite{PRL}, obtained through a non-local mean field approach. In Sec. \ref{sec:nonlocal} we show how to correctly apply the mean-field approach to obtain results consistent with the field-theoretical description. Sec. \ref{sec:conclusion} concludes the article with a summary of the results and an outlook to future research directions. 

\section{The Haldane insulator}\label{sec:back}

Lattice bosons with non-local interactions can form a zoo of different phases with local order parameters, including superfluids (SF), density-waves (DW), supersolids, and dimerized phases (See for example  Ref.s~\onlinecite{Jacksh,Zoller, Pupillo}). Traditionally, the only phases without a local parameter were Mott insulators at integer filling. These phases are topologically trivial because their ground states can be adiabatically deformed to site-factorizable states with an exact (integer) number of atoms per site (~$|11111...\rangle,~ |22222...\rangle,~...$ states )
In this sense, the Mott insulators (MI) of bosons are analogous to the trivial band insulators of fermions. In Ref.~\onlinecite{PRL}, we found that non-local (short-ranged) interactions can stabilize a non-trivial topological phase, the Haldane insulator (HI).

In our theoretical study, we considered the one-dimensional Bose-Hubbard model with extended interactions:
\bea H&=&\sum_i \left[-t (b\yd_i b\nd_{i+1}+{\rm H. c.})+{U\over 2} n_i(n_i-1)\right]\nn\\
&&+\sum_{i} \sum_{j>0} V_{j} n_i n_{i+j}\label{eq:eBH}\;. \eea
The first term $t$ describes the tunneling (``hopping'') between neighboring sites, the second term $U$ describes the effective interactions between atoms on the same site, and $V_{j}$ describes non-local interactions. This model applies to atoms and molecules with fixed electric or magnetic dipoles \cite{Lahaye,Pfau,Jin}, and to Rydberg atoms with induced dipoles \cite{Pupillo}. For simplicity, we will assume that only the nearest-neighbor term $V_{1}=V$ is non zero. This assumption is not essential for the present analysis, provided that $V_{j}$ decays fast enough.

\begin{figure}[t]
\centering\includegraphics[scale=1.0]{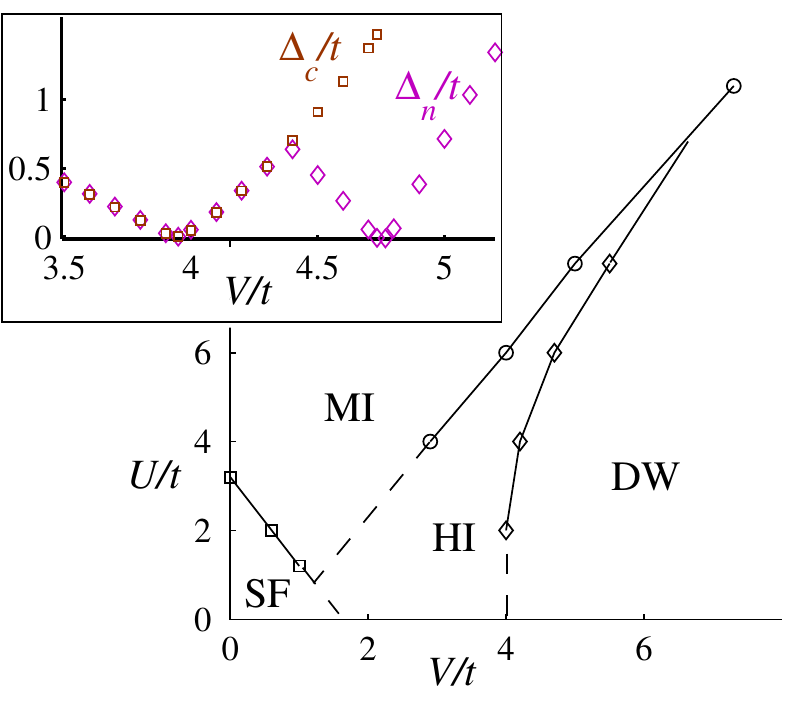}
\caption{Phase diagram of the extended Bose Hubbard model (\ref{eq:eBH}) at filling of $\bar n=1$ particles per site, as obtained by Density Matrix Renormalization Group (DMRG) calculations with $L_{\rm max}=256$ sites and $M_{\rm max}=250$ states per block. INSET: Excitation gaps $\D_c$ and $\D_n$ (defined in the text) along the path $U/t=6$. The phase transitions are located at the points in which the gap vanishes ($V/t\approx4$ and $V/t\approx4.75$).}
\label{fig:oldresults}
\end{figure}

At $V=0$ the Bose-Hubbard model describes a quantum phase transition \cite{Jacksh,amico00,Paredes,giamarchibosons,giamarchireview} between a one-dimensional superfluid (for small $U/t$) and a Mott insulator (for large $U/t$) . In the presence of a finite non-local repulsion $V>0$ (of the order of $U$) the Hamiltonian (\ref{eq:eBH})  describes a transition to a topological phase, the Haldane insulator \cite{PRL,PRB}. This transition was first discovered by numeric studies of the excitation gap, reproduced in the inset of Fig.\ref{fig:oldresults}.  This graph clearly shows the existence of three gapped phases, separated by two continuous phase transitions. The three gapped phases are the Mott insulator (MI), the Haldane insulator (HI), and the density-wave (DW) phases. This last phase is adiabatically connected to a classical state with an alternating number of bosons per site ($|202020...\rangle$), and spontaneously breaks the lattice translational invariance, modulo two sites ($Z_2$ symmetry).

The Haldane insulator is a symmetry-protected topological phase \cite{PRB,Pollmann,Pollmann2}. In analogy to the topological insulators of fermions (which are protected by time-reversal symmetry), the HI is protected by an anti-unitary discrete symmetry, the lattice-inversion symmetry. In the presence of this symmetry, the HI  cannot be adiabatically connected to a site-factorizable state. Remarkably, the lattice-inversion symmetry is trivially broken at the edges of the lattice, and therefore the Haldane insulator does not have protected edge states \footnote{As shown in Ref.s~\onlinecite{Pollmann,Pollmann2}, additional symmetries can protect the HI phase, and in particular a bosonic version of time-reversal invariance. If one of these symmetries is present, the system develops protected gapless edge states, analogous to those of the TI of fermions.}.

Because both the Mott insulator and the Haldane insulator do not break any local symmetry, they cannot be distinguished by any local order parameter. Exploiting an analogy to spin chains, we demonstrated that these two phases can nevertheless be characterized by two ``string'' order parameters \cite{PRB}. These order parameters are non-local in the sense that they consist of an infinite {\it product} of operators acting on different sites \cite{denNijs,Tasaki}.  Their experimental detection therefore requires the simultaneous measurement of all the sites in the lattice. Using new microscopic techniques for ultracold atoms \cite{Greiner-micro,Bloch-micro}, the string order of the Mott insulator has been recently measured \cite{Bloch-string}. 

The existence of a string order is however specific to one dimension and, to the moment, does not seem to extend to other topological phases \footnote{The decay of the string order parameter of the Mott insulator in two dimensions has been studied in Ref.~\onlinecite{rath13}} . In this article, we present an alternative method to characterize the topological nature of these phases. Our approach consists in probing the low-energy excitations of the different phases. It is therefore not specific to one-dimension and can be easily generalized to higher dimensions. In one dimension, both methods apply and could be used to complement and verify one-another.


\subsection*{Elementary excitations} As explained in the introduction, we propose here to study the topological properties of the Mott and Haldane phases by probing the low-energy excitations at the transition between the two phases. To characterize their nature, let us consider again the inset of Fig.\ref{fig:oldresults}. This graph  shows two independent curves. One curve (diamonds) represents the ``neutral gap''  $\Delta_n=E^{(1)}_{\delta N=0}-E^{(0)}_{\delta N=0}$, which is the gap to the first excited state with the same number of particles as in the ground state ($\delta N=0$). It was calculated by targeting the first excited state in the DMRG calculation. The second curve (squares) represents the ``particle-hole gap'' $2\Delta_c=E^{(0)}_{\delta N=1}+E^{(0)}_{\delta N=-1}-2E^{(0)}_{\delta N=0}$, which was calculated by adding or subtracting one particle ($\delta N=\pm1$) and computing the new ground state.

At the MI-HI phase transition the particle-hole gap closes, indicating that at this point the energy needed to create a particle or a hole (``the mass term'', in the field-theoretical terminology) becomes zero. At the HI-DW phase transition, on the other hand, the particle-hole gap remains open, indicating that the phase transition is due to a different type of excitations. We can easily guess the nature of these excitations by recalling that the HI-DW phase transition involves the spontaneous breaking of a local $Z_2$ symmetry, related to translational invariance modulo two lattice sites. In order to break this symmetry, the ground state has to mix with a phonon at the edge of the Brillouin zone (i. e. at momentum $k=\pi$), whose excitation gap must be closed at the phase transition.

\begin{figure}[t]
\centering
\includegraphics[scale=0.8]{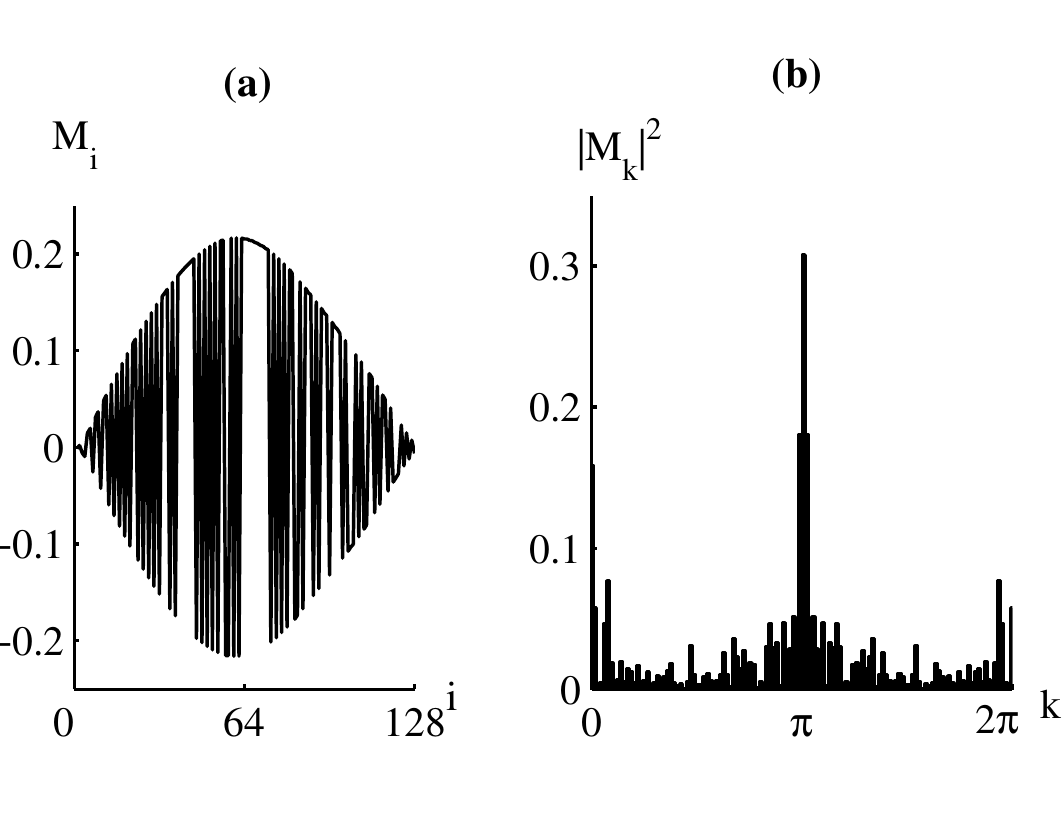}
\caption{Characterization of the lowest neutral excitation $\ket{\Psi^{(1)}_{\delta N =0}}$. {\bf (a)} Matrix element
of $M_i = \bra{\Psi^{(1)}_{\delta N =0}} \delta n_i \GS $ as function of site $i$. {\bf (b)} Square of the Fourier transform $\hat{M_k} = \sum_i \exp(i k x_i / L) M_i = \braket{\Psi^{(1)}_{\delta N =0}}{\psi^{\rm phonon}_{k}}$ as function of momentum $k$. The pronounced peak of $|M_k|^2$ for $k=\pi$ confirms that $\ket{\Psi^{(1)}_{\delta N =0}}$ has a large overlap with a phonon at the edge of the Brillouin zone}
\label{fig:deltafun}
\end{figure}

To verify this hypothesis, we compute the overlap between the lowest excited state in the vicinity of the HI-DW phase transition and a trial wavefunction for a phonon at momentum $k$:
\be \ket{\psi^{\rm phonon}_k} \equiv \sum_j e^{i k x_j} \delta n_j \GS\label{eq:phonon}.\ee 
Here $\GS$ is the ground state wavefunction as given by the DMRG algorithm and $\delta n_j \equiv n_j - \bar{n}$ the difference between the local occupation and the average occupation $\bar{n}$. As shown in Fig. \ref{fig:deltafun}, the numeric result confirms the existence of a large overlap with a phonon at momentum $k=\pi$, supporting our assumption.


\section{Critical theory of the phase transitions}\label{sec:methods}

The properties of any system near a continuous phase transitions do not dependent on the details of the microscopic model and are determined by a universal critical theory. The critical theories of the MI-HI and HI-DW phase transitions were derived in Ref.~\onlinecite{PRB} using a field-theoretical approach, the ``bosonization of bosons'' \cite{giamarchibosons, giamarchireview}. In what follows, we outline the properties of these theories, which will be used to compute the dynamic response of the system in Sec.~\ref{sec:results}.

\subsection*{MI-HI phase transition} The critical theory of the MI-HI phase transition is described by the sine-Gordon model
\be H = \int dx ~\left[\frac{K}2(\partial_x \theta)^2 + \frac{1}{2K}(\partial_x\phi)^2 - g \cos(2\phi)\right]\label{eq:SG}.\ee
Here the field $\phi(x)$ measures the displacement of the particles with respect to a fixed lattice and is related to the particles' density by $n(x)\approx n_0 - (1/\pi)\partial_x\phi$,
where $n_0=\bar n/a$, $\bar n$ is the average number of particles per site, and $a$ is the lattice constant. The field $\theta(x)$ is canonically conjugate to $n(x)$ and corresponds to the coherent phase of the bosons. 

The quadratic part of the Hamiltonian (\ref{eq:SG}) simply describes linear phonons, i.e. the Goldstone mode associated with translational invariance, in one dimension. The unitless prefactor $K$ is non-universal and it varies along the MI-HI phase boundary. A renormalization group (RG) analysis \cite{sineGordon} shows that, in order for a direct MI-HI phase transition to occur, $K$ must be bounded between $1/2$ and $2$. For $K>2$, the MI and HI phases would be separated by a superfluid phase, and for $K<1/2$, by a third gapped phase. 

The cosine term in Eq.~(\ref{eq:SG}) describes the effects of the periodic potential (lattice). This term pins the field $\phi(x)$ to a fixed value, modulo $\pi$. The MI-HI phase transition corresponds to a change in the sign of the prefactor. In the Mott insulator $g>0$ and the field $\phi$ is pinned to zero, indicating that the particles are centered at the lattice sites. In the Haldane insulator $g<0$ and $\phi$ is pinned to $\pi/2$. The particles are centered at the lattice bonds, and live in a coherent superposition of two (or more) neighboring sites. A weak-coupling analysis leads to $g \approx \rho_0(U-V)$, indicating that the MI-HI phase transition should be located approximately at $U=V$.

The low-energy excitations of the MI-HI phase transition correspond to jumps of $\pm \pi$ in the field $\phi(x)$, termed ``kinks", or ``solitons''.  Because the density of particles is proportional to the derivative of $\phi$, each kink involves the addition, or subtraction, of one particle. Using a perturbative RG approach it is possible to show that each kink increases the energy of the state by a charge gap $\D_c \sim g^{1/(2-K)}$. In addition, depending on the value of $K$, the kinks may either repel (for $K>1$), or attract (for $K<1$). In the former case, the neutral excitation gap is given by twice the energy cost of a kink, $\D_n = 2\D_c$. In the latter case, particles and holes form bound states (``breathers"). Their total mass is lowered by the binding energy of the pair. In what follows we consider only $1<K<2$, corresponding to intermediate values of the lattice depth.

The sine-Gordon model (\ref{eq:SG}) is a useful representation of the critical theory of the MI-HI phase transition because it is exactly solvable. In particular, the matrix elements between the ground state and any excited state can be obtained following the ``form factor" approach introduced by Smirnov  \cite{Smirnov}. This approach relies on the exact expression for the scattering-matrix between kinks and breathers.

\subsection*{HI-DW phase transition}
The transition from the Haldane insulator (HI) to the density-wave (DW) phase breaks a local $Z_2$ symmetry. In Ref.~\onlinecite{PRB} we showed that, indeed, this transition belongs to the universality class of the Ising transition in 1+1 dimensions. The critical theory of this transition is captured, for instance, by the spin-1/2 transverse-field Ising chain
\be H = \sum_i \left[-\s^z_i \s^z_{i+1} + \lambda \s^x_i \right] \label{eq:Ising}.\ee
Here $\s^z_i$ and $\s^x_i$ are Pauli matrices. The model (\ref{eq:Ising}) shows a quantum phase transition at $\lambda=1$ between a paramagnet, with zero magnetization $\av{\s^z_i}=0$, and a ferromagnet, with $\av{\s^z_i}\neq0$. The ferromagnet breaks the $Z_2$ symmetry, related to the rotation of $\pi$ around the $z$-axis. The transverse-field Ising model (\ref{eq:Ising}) has an exact solution in terms of free fermions, based on the Jordan-Wigner transformation \cite{JordanWigner,Sachdev}. This mapping shows that the excitation gap vanishes linearly and symmetrically around the phase transition: \be \D = |1-\lambda|\;. \ee

The exact solution also shows that the nature of the low-energy excitations changes dramatically across the phase transition. In the paramagnetic phase, the low-energy excitations are local spin-flips, created by the operator $\s^z_i$. In the ferromagnetic phase, on the other hand, the elementary excitations are domain walls between spin-up and spin-down regions.
 
 In the original Bose-Hubbard model the paramagnetic and ferromagnetic phases correspond respectively to the Haldane insulator and the density-wave phase. The spin-up and spin-down states correspond to the two possible configurations of the density-wave ($|202020...\rangle$ and $|020202...\rangle$), leading to the mapping 
\be  \s^z_i ~~\leftrightarrow~~ (-1)^i (n_i - \bar n) \label{eq:isingorder}\;.\ee
Eq.~(\ref{eq:isingorder}) indicated that the low-energy excitations of the Haldane insulator are density fluctuations (phonons) with momentum $\pi$. In contrast, the low-energy excitations of the density-wave phase are non-local domain walls. 

\section{Dynamic probes: absorption spectra}\label{sec:results}

We now apply the above field-theoretical description to compute the dynamic response of the system in the vicinity of the MI-HI and HI-DW phase transitions. In particular, we focus on two specific perturbations: lattice modulations and Bragg spectroscopy at momentum $k=\pi$. As we will see, these perturbations are easily realizable with ultra-cold atoms and best capture the differences between the two transitions. The results obtained by this method are summarized in Fig.s~ \ref{fig:lattice} and \ref{fig:bragg}.

In an idealized version of the proposed experiment the system is initially prepared in the ground state, a periodic perturbation is applied with frequency $\w$ for a given time $t_{\rm pert}\gg1/\w$, and finally the amount of absorbed energy $dE$ is computed, by measuring the increase in temperature of the system. The absorption spectrum is defined as the ratio between the absorbed energy and the perturbation time $t_{\rm pert}$. According to Fermi golden rule, this ratio is independent on $t_{\rm pert}$, and equals to
\be S(\w) = \sum_a \left| \bra{\psi_\a} T \GS \right|^2 \left(\ve_\a-\ve_{\rm gs}\right) \delta(\w+\ve_{\rm gs}-\ve_\a).\label{eq:Fermi}\ee 
Here $\GS$ and $\ket{\psi_\a}$ are, respectively, the ground- and excited-states of the unperturbed Hamiltonian, with energies $\e_{\rm gs}$ and $\ve_\a$ (we work in units where $\hbar =1$). The operator $T$ models the perturbation and will be derived below for two specific cases, denoted by $T_{\rm hop}$ and $T_{k=\pi}$.

The first perturbation consists of small modulations of the lattice strength. It  mainly influences the tunneling matrix element between two neighboring sites and corresponds to the operator
\be T_{\rm hop} \equiv \delta t \sum_i \left[b\yd_i b\nd_{i+1} + {\rm H. c.}\right]\;.\label{eq:Thop}\ee
The second perturbation corresponds to the addition of a superlattice with a given momentum $k$, which locally modifies both the tunneling and the chemical potential. In the case of $k=\pi$, the tunneling  is modulated uniformly along the lattice, while the chemical potentials of odd and even sites are shifted in opposite directions. The correspondent operator is the sum of (\ref{eq:Thop}) and
\be T_{k=\pi} \equiv  \delta \mu \sum_i (-1)^i \delta n_i \label{eq:Bragg}\;,\ee
where $\delta n_i = b_i\yd b_i - \bar{n}$. In the following we will assume that the Bragg spectroscopy is performed by modulating both the lattice and the superlattice in such a way that only $T_{k=\pi}$ is applied.


\subsection*{MI-HI phase transition} The critical theory of the MI-HI phase transition can be expressed  in terms of the sine-Gordon model (\ref{eq:SG}). In this language, the modulation of the lattice corresponds to \cite{PRB}
\be T_{\rm hop}\approx \d t \int dx~ \cos(2\phi) \;.\ee
This term creates kink-anti-kink pairs, corresponding to particle-hole excitations. The total energy of the pair is given by the sum of two positive terms: the chemical potential, $\D_n = 2\D_c$, and the kinetic energy of the pair. As a consequence, the resulting absorption spectrum $S_{\rm hop}(\w)$ is non zero only for frequencies $\w>2\D_c$. The relevant matrix elements can be computed using the form factor approach \cite{Smirnov, Gogolin, Lukyanov2001}, leading to the formulas summarized in Appendix \ref{app:form}.  Close to the lowest edge of the continuum  $0< \w-2\D_c \ll \D_c$ the spectrum follows the power law
\be S_{\rm hop}(\w) \sim (\w-2\D_c)^{2K} \;.\label{eq:Shop1}\ee

Bragg spectroscopy, on the other hand, does not couple to the low-energy excitations of the MI-HI phase transitions. The critical excitations of this phase transition are particles and holes with zero momentum, while $T_{k=\pi}$ couples only to excitations with total momentum  $k=\pi$. Thus, Bragg spectroscopy at momentum $k=\pi$ does not show any particular low-frequency feature at the MI-HI phase transition, despite the closing of the gap. This observation will allow us to clearly distinguish between this transition and the HI-DW transition to be considered below.

\subsection*{HI-DW phase transition} The critical theory of the HI-DW phase transition is captured by the transverse-field Ising model (\ref{eq:Ising}). The operator $T_{\rm hop}$ corresponds to the term of the Hamiltonian (\ref{eq:eBH}) which stabilizes the Haldane insulator, and therefore maps to the $\sum_i \sigma^x_i$ term of Eq.~(\ref{eq:Ising}) which stabilizes the paramagnetic phase. In Appendix \ref{app:jordan}, we show how to compute the response of this term by mapping Eq.~(\ref{eq:Ising}) to a theory of free fermions, through the subsequent application of the Jordan-Wigner and Bogoliubov transformations. The resulting absorption spectrum is:
\be S_{\rm hop}(\w) = d t^2 \frac{\sqrt{\omega^2-(2\D_n)^2}}{\omega}\Theta(\omega-2\D_n)\;.\label{eq:Shop2}\ee
Here $\Theta$ is the Heaviside theta function, and $\D_n=|1-\lambda|$ is the neutral excitation gap. As explained in Sec.\ref{sec:back}, the gap $\D_n$ separates the ground state from a phonon at momentum $k=\pi$. Lattice modulations correspond to a translational invariant operator and can couple only to excitations with zero total momentum, such as, in this case, a pair of counter-propagating phonons. The absorption spectrum (\ref{eq:Shop2}) shows a broad feature, corresponding to the continuum of two-particle excitations. This result is in contrast with the analysis of Ref.~\onlinecite{PRL}, which predicted the appearance of a sharp peak in the response to lattice modulations. As we will see in Sec.~\ref{sec:nonlocal} the mean-field approach can be fixed by a more accurate characterization of the elementary excitations.
 
The operator corresponding to Bragg spectroscopy (\ref{eq:Bragg}) is a modulation of the local operator $(-1)^i\delta n_i$, which corresponds to the order parameter of the symmetry-braking phase and, in the effective Ising model, maps to the local magnetization $\sigma^x_i$  (see Eq.~(\ref{eq:isingorder})). In the paramagnetic phase (Haldane insulator), this operator couples directly to the low-energy excitations of the phase: spin-flips at zero-momentum. The corresponding matrix element can be computed through conformal field theory (CFT) methods \cite{Polyakov}, leading to
\be S_{k=\pi}(\w) \approx |\D|^{2\eta-1} \delta (\omega-\Delta_n).\label{eq:Spi2}\ee 
Here $\eta=1/8$ is the anomalous dimension of the operator $\sigma^x$.  The operator $T_{k=\pi}$ can additionally couple to any state with an odd number of spin-flips. Therefore we expect a continuum of three-particles excitations in $S_{k=\pi}$ for $\omega > 3|\Delta_n|$.

In the DW phase (ferromagnet) the elementary excitations are domain walls in the ferromagnetic order parameter $\av{\sigma^z}$. The operator $\sigma^x$ is local and  can excite only an even number of domain walls. Thus, in this phase we expect a broad continuum, whose lowest edge is located at $\omega = 2|\Delta_n|$.  

\begin{figure}[t] \centering

\includegraphics[scale=0.7]{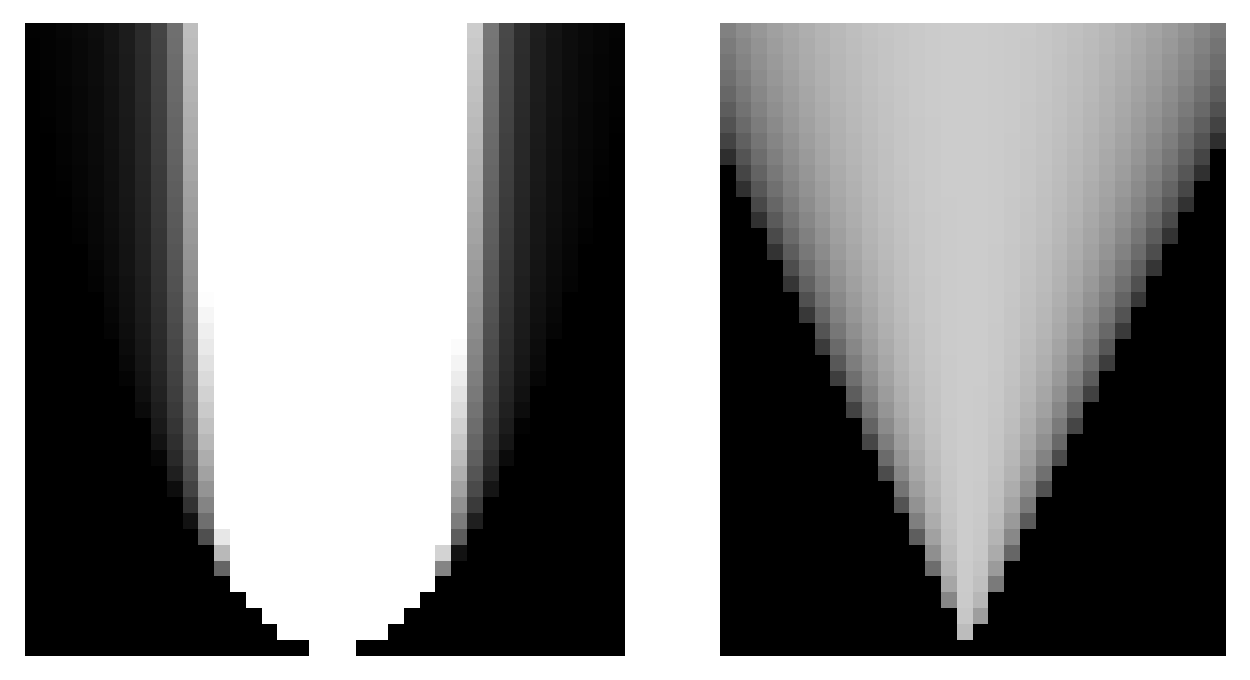}
\caption{Energy absorption due to lattice modulations $S_{\rm hop}(\w)$, according to the effective field-theoretical description. The horizontal axis corresponds to generic paths crossing the MI-HI (left panel) and the HI-DW  (right panel) phase transitions. The vertical axis corresponds to the frequency $\w$. Both axes are in arbitrary units: the field-theoretical approach describes the universal shape of the spectra around the phase transitions, but does not deliver quantitative information about their position. The value of $K$ at the MI-HI phase transition (\ref{eq:SG}) is arbitrarily set to $K=1.5$. Lattice modulations couple to pairs of low-lying excitations near both transitions, leading to broad response spectra.}
\label{fig:lattice}
\vspace{1cm}
\includegraphics[scale=0.7]{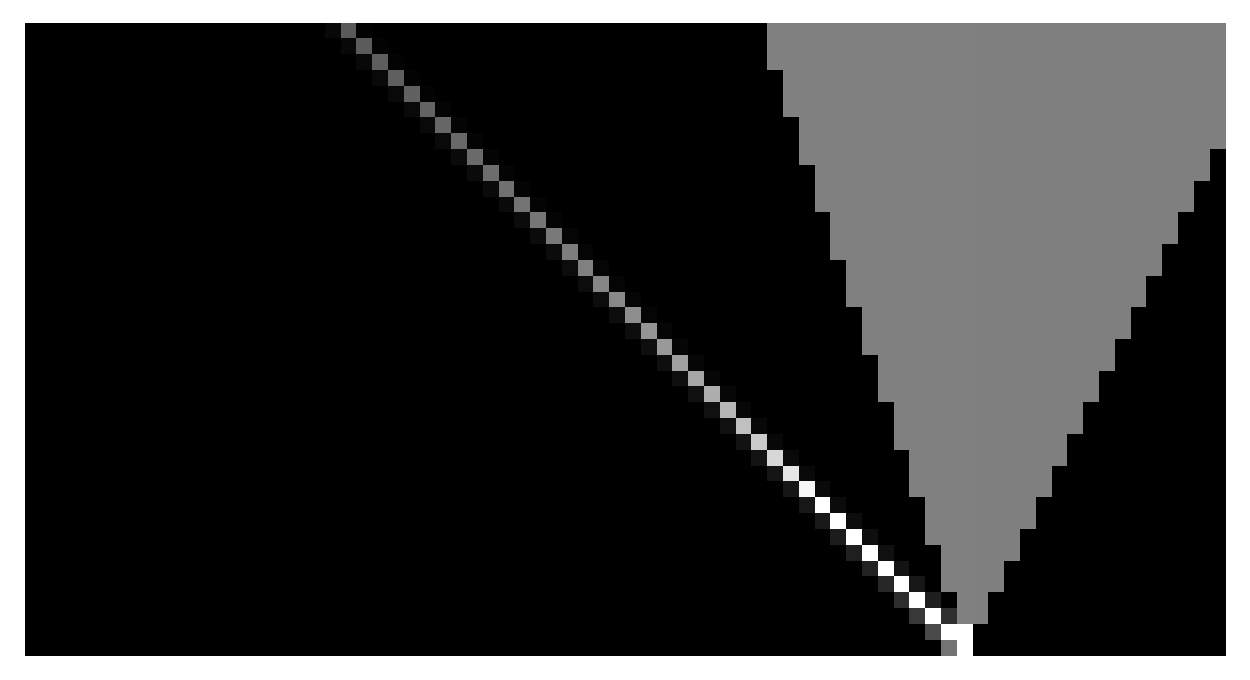}
\caption{Energy absorption due to Bragg spectroscopy across the HI-DW phase transition, for the same parameters as in Fig.~\ref{fig:lattice}. Bragg spectroscopy at momentum  $k=\pi$ couples to a single phonon whose gap closes at the HI-DW phase transition}
\label{fig:bragg}
\end{figure}

Fig.s \ref{fig:lattice} and \ref{fig:bragg} summarize graphically our findings for the absorption spectrum of lattice modulation and Bragg spectroscopy at the MI-HI and HI-DW phase transitions. The extended bright areas indicate the excitation of pairs or triplets of elementary excitations. The narrow absorption peak in Bragg spectroscopy is the signature of the single phonon at momentum $k=\pi$, needed to break the translational invariance

\section{Non-local mean field}
\label{sec:nonlocal}
We now present an alternative method to compute the dynamic response of the Haldane insulator, based on a non-local mean field introduced by Kennedy and Tasaki  \cite{KennedyTasaki}. This approach is valid deep in the gapped phase, and it complements the field-theoretical approach presented above, which is valid only in the vicinity of a phase transition. 

The first step of this method is an approximate mapping between the Haldane insulator and the Haldane gapped phase of spin-1 chains \cite{PRL}. The mapping can be made explicit by truncating the Hilbert space of the Bose system to three occupation states per site (for example, for a system with ${\bar n}=1$ particles per site, we keep only the occupation states $n=0,1,2$ for every site). This defines an effective spin-1 model with $S^z_i=n_i-{\bar n}$.  If we neglect particle-hole symmetry breaking terms, the analogous spin-1 model is the well-known antiferromagnetic XXZ model
\be H = \sum_i -t \left(S^+_i S^-_{i+1} + {\rm H. c.}\right) + V S^z_i S^z_{i+1} +\frac{U}{2} (S^z_i)^2\label{eq:XXZ}\;.\ee
This model supports three gapped phase, the ``large-U'', the Haldane, and the Neel phases, corresponding respectively to the MI, HI and DW phases.

Kennedy and Tasaki \cite{KennedyTasaki} introduced a non-local mean-field theory which captures the whole phase diagram of the XXZ chain. Their approach relies on a unitary transformation \cite{Oshikawa} that transforms the string operators into local spin operators:
\begin{align}
O^x_{string} &\equiv S^x_i \exp\left(i\pi\sum_{j>i} S^x_j \right) = U^{-1} S^x_j U = \tilde{S}^x_i \nn\;,\\
O^z_{string} &\equiv \exp\left(i\pi\sum_{j<i} {S}^z_j\right) {S}^z_i = U^{-1} S^z_j U =  \tilde{S}^z_i\label{eq:Ostring}\;.
\end{align}
The operator $U$ flips every second site with $S^z=\pm 1$, skipping those with $S^z=0$.  It is given explicitly by \cite{Oshikawa} 
\be U=\prod_{\forall j,k|j<k}e^{i\pi S^z_j S^x_k }\equiv\prod_{\forall j,k|j<k}
E_{j,k} \label{unit} \;.\ee


The unitary operator $U$ transforms the Hamiltonian (\ref{eq:XXZ}) into an unusual, but nevertheless local form 
\bea \tilde{H} = UHU &=&-2t\sum_{i}\left[
\tilde{S}^x_{i} \tilde{S}^x_{i+1} - \tilde{S}^y_{i} \exp(i \pi \tilde{S}^z_i + i \pi \tilde{S}^x_{i+1})
\tilde{S}^y_{i+1} \right]\nn\\&&- V\sum_{i}\left[ \tilde{S}^{z}_i \tilde{S}^{z}_{i+1}\right] + {U\over 2} \sum_i (\tilde{S}^{z}_i)^2\;.
\label{Htilde} \eea
The transformed Hamiltonian $\tilde{H}$ has an explicit $Z_2 \times Z_2$ symmetry, generated by $\pi$ rotations of the spins around the main axes. In the Haldane phase this symmetry is spontaneously broken and the two order parameters $\tilde{S}^x$ and $\tilde{S}^z$ acquire a non-zero mean value. These properties enable a variational mean-field approximation  \cite{KennedyTasaki} for the ground state. It consists in considering only site-factorizable wavefunctions of the form \be \ket{\Psi^\a_{\rm GS}} = \prod_{1<i<L} \ket{\psi_\a}_i \triangleq \ket{ \a \a ...\a \a}\;, \ee
where $\ket{\psi_\a} = \cos\t\ket{0} \pm \sin\t\ket{\pm 1}$.

\begin{figure}[h] \centering
\includegraphics[scale=0.75]{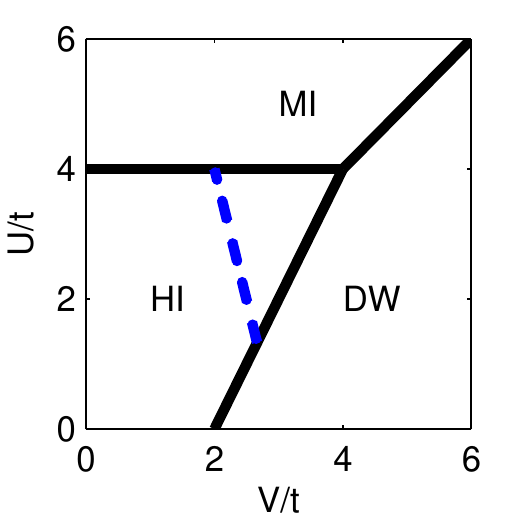}
\caption{Phase diagram (solid lines) of the effective spin-1 model (\ref{Htilde}): the three phases map to the Mott insulator (MI), the Haldane Insulator (HI) and the density-wave (DW) phase. The dashed line is one particular paths crossing the HI phase and used to plot Fig.s \ref{fig:lattice2} and \ref{fig:bragg2}.}
\label{phdiaUVJ}
\end{figure}

In the Haldane insulator the minimal variational energy is obtained for $\cos^2\t=(U-2V+4t)/(8t-2V)$ and the ground state is four fold degenerate. As shown in Table
\ref{4states} the four different states can be characterized by the signs of the two order parameters and reflect the broken $Z_2 \times Z_2$ symmetry. 
In the DW phase $\cos\t=0$ and the ground state is doubly degenerate. In the
MI phase $\cos\t=1$ and the ground state is  non-degenerate. The resulting phase boundaries, first obtained in
 Ref.~\onlinecite{KennedyTasaki}, are given in Fig. \ref{phdiaUVJ}.

\begin{table}[h]
\centering
\begin{tabular}{|c|c|c|}
\hline
 {\rm state $\backslash$ order parameter} & ${\rm sign}\av{\tilde{S}^x}$ & ${\rm sign}\av{\tilde{S}^z}$\\
\hline
$ \ket{\psi_1} = \ct|0\ra + \st |+1\ra$ &+&+ \\
\hline
$ \ket{\psi_2}= \ct |0\ra - \st |+1\ra$ &-&+\\
\hline
$ \ket{\psi_3}= \ct|0\ra + \st |-1\ra$ &+&-\\
\hline
$\ket{\psi_4} = \ct|0\ra - \st |-1\ra$ &-&-\\
\hline
\end{tabular}
\caption{Four mean-field ground-states of the transformed Hamiltonian (\ref{Htilde}) and the respective signs of $\av{\tilde{S}^x}$ and $\av{\tilde{S}^z}$}
\label{4states}\end{table}

\subsection*{Variational wavefunctions of the excitations}
The non-local mean field allows to describe the elementary excitations of the different phases in terms of domain walls (``kinks'') of the two order parameters \cite{Solyom}. A domain wall localized at site $i$ may be represented by the site factorizable wavefunction
\bea \ket{\Psi^{\a,{\b}}_i} &=& \prod_{1<j<i} \ket{\psi_\a}_j \prod_{i+1<j'<L} \ket{\psi_{\b}}_j  \\&\triangleq& \ket{\a\a...\a\a\;\b\b...\b\b}_i\;. \eea
This wavefunction can be symmetrized with respect to lattice translations to obtain a variational ansatz for the elementary excitations:
\be \ket{\Psi^{\a,\b}_k} =\sum_j e^{i k x_j} \ket{\Psi^{\a,\b}_j} \triangleq \ket{\a\a...\a\a\;\b\b...\b\b}_k\;. \ee
In the Haldane insulator, the domain walls can flip either one of the hidden $Z_2$ symmetries or both, giving rise to three different excitations. Note that these excitations do not coincide with those predicted in Ref.~\onlinecite{PRL}. As we will see, the present approach correctly reproduces the findings of the field-theoretical approach.

A kink that flips only $\tilde{S}^x$ is described by the wavefunction
$\ket{\Psi^z_k}=\ket{\Psi^{1,2}_k}\triangleq \ket{11...1122...22}_k$ (see Table \ref{4states}) and is created by the untransformed 
operator $S^z_k$. 
In terms of the original Bose-Hubbard model, this ``neutral'' excitation corresponds to a phonon (\ref{eq:phonon}) with momentum $k=\pi$.
In contrast, the operators $S^+_k$ and $S^-_k$ change the total number of particles by one.  The resulting excitations $\ket{\Psi^{\pm}_k} =
\ket{\Psi^{x}_k} \pm \ket{\Psi^{y}_k}  = \ket{\Psi^{1,3}_k} \pm \ket{\Psi^{2,3}_k} \equiv \ket{11...1133...33}_k +
\ket{22...2233...33}_k$ correspond to a superposition of two types of kinks: 
$\ket{11...1133...33}_k$, which is a kink only in $\tilde{S}^z$, and $\ket{22...2233...33}_k$, which is a kink in both
$\tilde{S}^x$ and $\tilde{S}^z$. In the original Bose-Hubbard model these ``charged'' excitations correspond to the creation or annihilation of one
boson with momentum $k$.

The energy of the three collective excitations can be calculated variationally: $E^\a = \la\Psi^\a_k|\tilde{H}|\Psi^\a_k\ra
- \la\Psi_{GS}|\tilde{H}|\Psi_{GS}\ra$, 
leading to 
\begin{align}
\D_n &= E^z_k = 2t (1 + \cos^2 2\t - 2\cos 2\t \cos k) \label{Ez}\;,\\
\D_c &= E^x_k = E^y_k\nn \\
&= 2 V {1-\cos^2\t \over 1+\cos^2\t}(1+\cos^4\t + 2\cos^2\t \cos k) \label{Ex}\;.\end{align}
We find that: (i) at the transition to the Mott insulator ($\theta\rar0$), the energies of both excitations vanish at momentum $k=0$; (ii) at the transition to the DW phase ($\theta\rar\pi/2$), only the energy of the neutral mode (\ref{Ez}) vanishes at wavevector $k=\pi$. Both observations are in agreement with the numerical and field-theoretical approaches presented above\footnote{Note that at the SU(2) symmetric Heisenberg point ($U=0,\;V=t\Rightarrow\cos^2\t=1/3$) the three excitations are
degenerate: they correspond to an SU(2) magnon triplet.}.


\subsection*{Absorption spectrum: selection rules} Having described the elementary excitations of the Haldane insulator, we can now calculate the absorption spectra $S(\w)$.  For this task, we should compute the matrix element between the different excitations and the perturbation operators defined in Eq.~(\ref{eq:Thop}) and (\ref{eq:Bragg}).
Unfortunately, these calculations are extremely complicated, due to the unknown interactions between the kinks. Nevertheless, we can formulate simple selection rules that determine when these matrix elements can be finite and when they must vanish.

To establish the selection rules, we classify all excited states in four topological sectors, defined by the parity of the number of kinks in the $x$ and $z$ string order parameters, denoted by $\mathcal{N}^x$ and $\mathcal{N}^z$. Because the string order parameters of the original model correspond to the local spin operators of the transformed model (Eq.~(\ref{eq:Ostring})), we can define the Boolean numbers $\mathcal{N}^x$ and $\mathcal{N}^z$ according to
\be \mathcal{N}^\a = \lim_{j\to\infty}{\rm sign}\av{\tilde{S}^\a_{-j}\tilde{S}^\a_{+j}}. \ee

The perturbations (\ref{eq:Thop})  and (\ref{eq:Bragg}) can connect the ground state only to excitations belonging to a well-defined topological sector. In particular, the operator corresponding to lattice modulations, $T_{\rm hop}$, remains local after the unitary transformation (\ref{unit})
\be UT_{\rm hop}U^{-1} = \sum_i \tilde{S}^x_{j} \tilde{S}^x_{j+1} - \tilde{S}^y_{j} \exp(i \pi \tilde{S}^z_j + i \pi \tilde{S}^x_{j+1}) \tilde{S}^y_{j+1}\;.\ee
As a consequence, $T_{\rm hop}$ cannot excite states with an odd number of domain walls, and
\be \bra{\psi_\a} T_{\rm hop} \GS \neq 0~~ {\rm only~if}~~ (\mathcal{N}^x,\mathcal{N}^z)_\a=(+,+)\;. \ee
Bragg spectroscopy transforms as
\be U T_{k} U^{-1} = \sum_i  e^{i k x_i} \tilde{S}^z_i \prod_{j<i} \exp(i\pi \tilde{S}^z_j)\;. \ee
Noting that $\exp(i\pi \tilde{S}^z_i)$ flips $\tilde{S}^x_i \to -\tilde{S}^x_i$, we deduce that $T_{k=\pi}$ flips $\mathcal{N}^x \to -\mathcal{N}^x$. Consequently $T_{k=\pi} \GS$ overlaps only with states containing an odd number of kinks of the $x$ string order parameter and
\be \bra{\psi_\a} T_{k=\pi} \GS \neq 0~~ {\rm only~if}~~ (\mathcal{N}^x,\mathcal{N}^z)_\a=(-,+)\;. \ee
Using these selection rules, we conclude that lattice modulations can excite only pairs of charged or neutral excitations and therefore its absorption spectrum is dominated by two particle continua. Bragg spectroscopy $T_{k=\pi}$ can couple to a neutral excitation at momentum $k=\pi$: we therefore expect a sharp resonance in the absorption spectrum, whose energy goes to zero at the HI-DW phase transition.

\begin{figure}[t]

\includegraphics[scale=0.7]{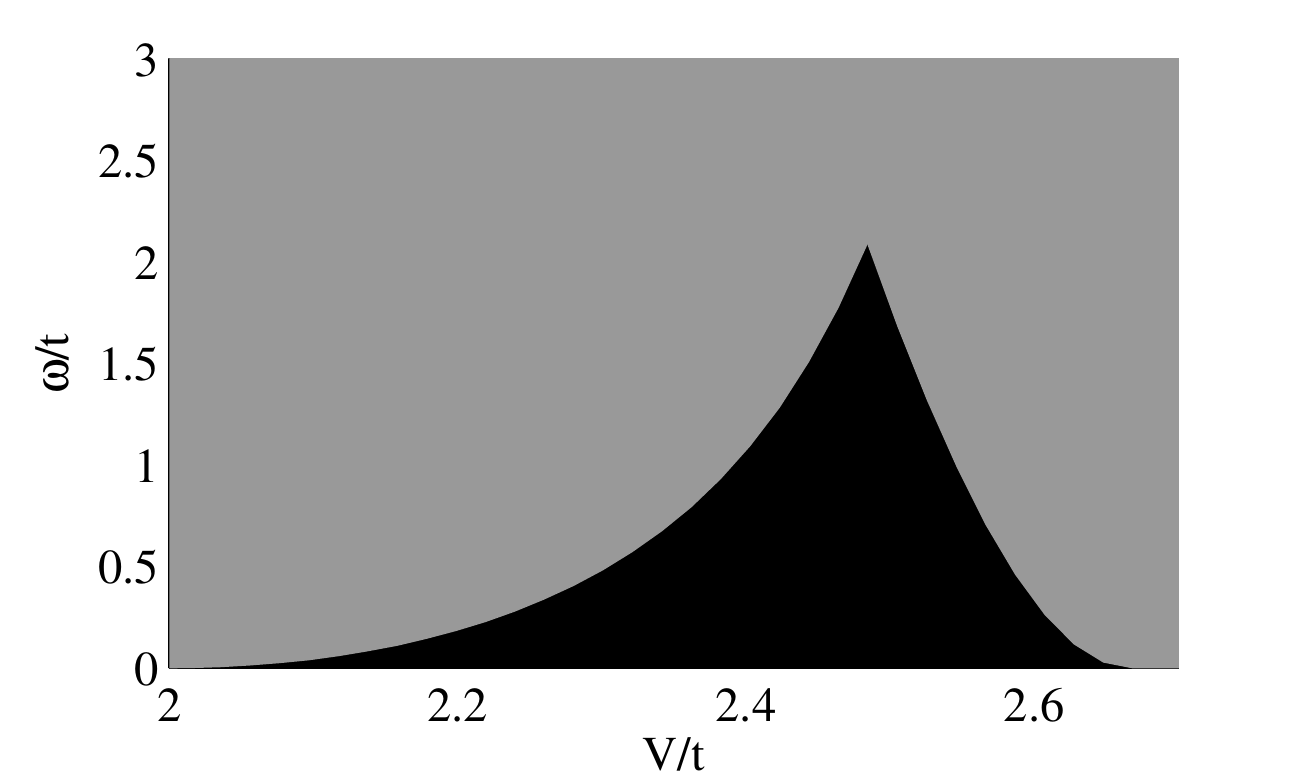}
\caption{Energy absorption due to lattice modulations $S_{\rm hop}(\w)$, according to the non-local mean-field description, along the dotted line in Fig. \ref{phdiaUVJ}.  Lattice modulations couple to both pairs of charged kinks (left gray area) and pairs of neutral kinks (right gray area).}
\label{fig:lattice2}

\includegraphics[scale=0.7]{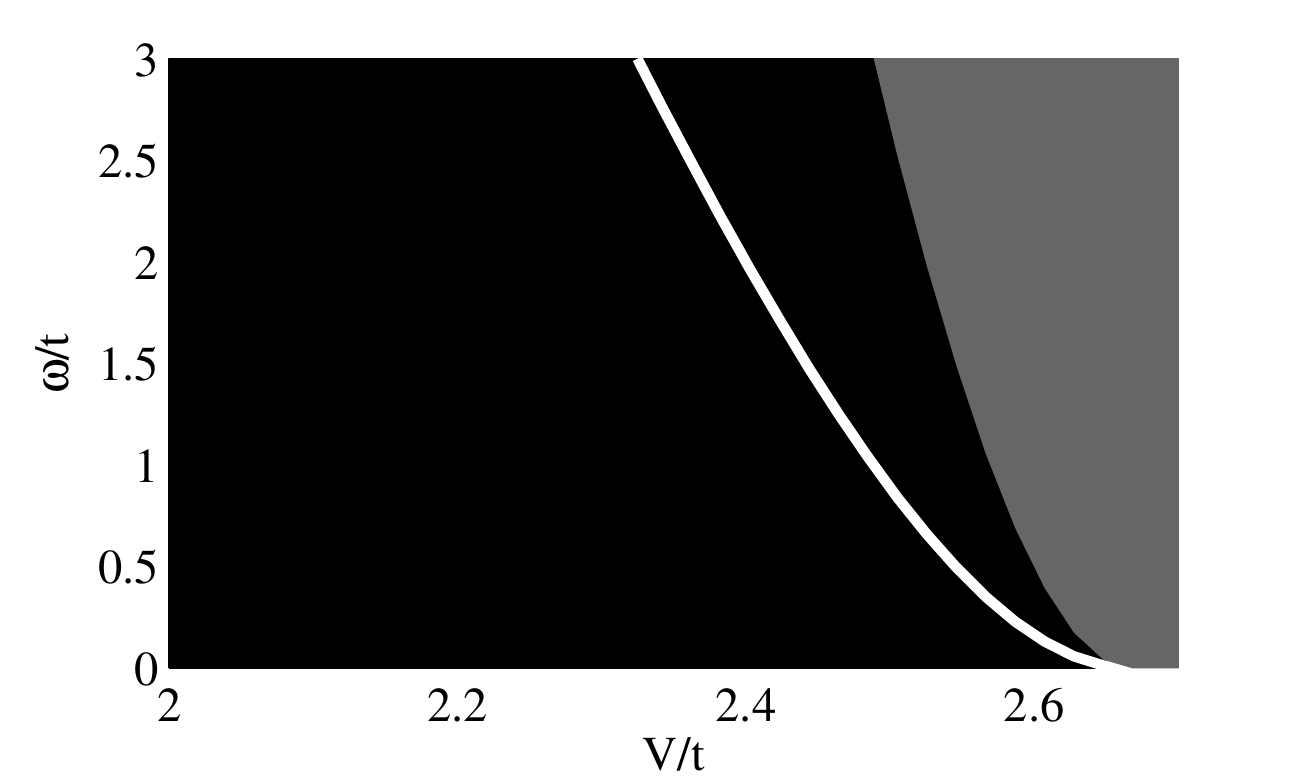}
\caption{Energy absorption due to Bragg spectroscopy, $S_{k=\pi}(\w)$, according to the non-local mean-field description, along the dotted line in Fig. \ref{phdiaUVJ}. Bragg spectroscopy at $k=\pi$ couples to single neutral kinks (white line), to pairs of charged kinks (gray area)}
{\label{fig:bragg2}}

\end{figure}

The resulting absorption spectra are shown in Fig. \ref{fig:lattice2} and \ref{fig:bragg2}. We find an excellent agreement with the results obtained from the critical theories shown in Fig. \ref{fig:lattice} and \ref{fig:bragg}. As expected, the mean-field approach does not capture the precise shape of the excitation gap in the vicinity of the transition. It does however provide a reliable method to quantitatively predict the position of the peaks and continua, deep in the Haldane insulator, starting from the microscopic  Hamiltonian (\ref{eq:eBH}).

\section{Conclusion}\label{sec:conclusion}

In this article we described a dynamical method to characterize the phase transitions between three insulating phases of interacting lattice bosons in one dimension: the Mott insulator (MI), the Haldane insulator (HI), and the density-wave (DW) phases. We considered the response of the system to two different perturbations, namely lattice modulations and Bragg spectroscopy at momentum $k=\pi$. Measuring the energy absorption rate of these perturbations, it is possible to characterize the critical excitations of the different transitions.

At the MI-HI phase transition, the low energy excitations are particles and holes at zero momentum. They appear as a broad peak in the response to lattice modulations, and they do not couple to Bragg spectroscopy at $k=\pi$. This observation indicates that the phase transition between the Mott insulator and the Haldane insulator is of topological origin: the critical excitations at a topological phase transition trivially commute with all the symmetries of the problem, i. e. translational invariance and particle conservation.

At the HI-DW phase transition, the low energy excitations are phonons at the edge of the Brillouin zone. These excitations appear as a narrow peak in the response to Bragg spectroscopy at $k=\pi$. They hold a non-trivial quantum number (momentum), showing that the HI-DW phase transition is not topological, and spontaneously breaks a local symmetry (translational invariance).

These qualitative features are confirmed by the field-theoretical analysis of the phase transitions. More quantitative predictions are obtained using a {\it non-local} mean-field approach. In contrast to the analysis of Ref.~\onlinecite{PRL}, the present approach correctly reproduces the shape of the spectra predicted by the field-theoretical calculations, at both the MI-HI and HI-DW phase transitions. The non-local mean-field can be therefore used to interpolate between the two phase transitions, and is expected to provide reliable results deep in the topological phase (HI), where the field theories cease to apply. This method does not rely on effective parameters of the continuum theory, which are in general not known, but rather applies directly to the microscopic Bose-Hubbard model.

The techniques developed here can be extended to other types of dynamic probes, such as Bragg spectroscopy at a general momentum $k$, local perturbations, small quenches \cite{deGrandi}, as well as to quantum phase transitions of other systems. Among those, we would like to briefly discuss quantum systems of interacting fermions in one dimension. The extended Bose-Hubbard model considered here can be approximately mapped into a fermionic problem, by splitting the single chain at integer filling into two chains at half-integer filling, and then applying the Jordan-Wigner transformation. This transformation maps the local occupation $n_i$ and the nearest neighbor tunneling $b\yd_i b_{i+1}$ of the bosons into the corresponding operators of the fermions. As a consequence, we expect the absorption spectrum of fermions to be identical to the one of bosons. The topological phase transition between the Mott insulator and the Haldane insulator maps into a phase transition between two topological phases of interacting fermions in one dimension, belonging to the more general $Z_8$ group of Ref.~\onlinecite{kitaev}. (The fermionic model possesses additional symmetries which map into local symmetries of the bosonic model.) It would be interesting to extend this result to the transitions to the other six phases.

Field-theoretical techniques should allow to extend the present approach to topological phases of fermions and bosons in higher dimensions. Recently, field theories describing gapped phases of fermions \cite{zhang11} and bosons \cite{wen12} have been found. However, less is known about the critical theory of the quantum phase transitions between the different topological phases. In the case of topological band insulators, it is natural to expect these critical theories to be non-interacting (Gaussian), in analogy to the present case. In general it should be possible to compute the response to physical probes by expressing the corresponding perturbations in terms of the critical theories. One should than be able to experimentally probe that the critical excitations of the topological phase transition respect all the symmetries of the systems.

\section{Acknowledgment} We wish to thank E. Berg, T. Giamarchi, S. Huber, S. Kivelson for many useful discussions. Part of the results presented in this work has been submitted to the Feinberg Graduate School, as a chapter of the Ph.D. thesis of the author, under the supervision of E. Altman.

\appendix

\section{Form-factor approach}
\label{app:form}
According to the form-factor approach \cite{Smirnov, Gogolin, Lukyanov2001}, the spectrum of lattice modulations is given by:
\be S_{\rm hop}(\w) = \frac{1}{\w \sqrt{\w^2 - 4 \D_c^2}} \left| ~ F_{\rm cos}\left[\theta(\w,g)\right]~\right|^2\;,\ee
where $\D_c$ is the charge gap (''soliton mass'') and
\begin{align} 
\theta(\w,g) &= 2 \log\left(\frac{\w}{2\D} + \sqrt{\frac{\w^2}{4\D_c^2}-1}\right)\;, \\
F_{\rm cos}(\theta) &= \frac{\cosh(\theta/2)}{\sinh\left[\frac{\pi}{2\gamma}(\theta-i \pi)\right]} F(\theta)\;, \\
F(\theta) &= \sinh(\theta/2) \exp\left(\int_0^\infty dx~K(\theta,x)\right)\;, \\ 
K(\theta,x) &= \frac{\left(\sin\left[x (\theta+i\pi)/2\right]\right)^2 \sinh\left[(\pi-\gamma) x/2\right]}{x \sinh(\pi x) \cosh(\pi x/2) \sinh(\gamma x/2)}\;,\\
\gamma &= \frac{\pi K}{2-K}\;.
\end{align}

\section{Jordan-Wigner and Bogoliubov transformations}
\label{app:jordan}

In this appendix we  compute the absorption spectrum of $T_{\rm hop}=\sum_i S^z_i$ through the subsequent application of Jordan-Wigner and Bogoliubov transformation. Applying Jordan-Wigner transformation to (\ref{eq:Ising}) we obtain:
\bea
H_{\rm Ising} &=& \sum_i n_i - \lambda (\psi\yd_i + \psi_i)(\psi\yd_{i+1} + \psi_{i+1})\label{eq:HJW}\\
&=&\sum_k (1-\lambda\cos(k))\psi_k\yd\psi_k + \lambda\sin(k)\psi_k\psi_k + {\rm H. c.}\nn 
\eea
Here $u=2 \lambda a$ and $\D = 2(\lambda-1)$. We now diagonalize (\ref{eq:HJW}) by the Bogoliubov transform: $\gamma_k = \cos(\theta_k/2)\psi_k -i\sin(\theta_k/2)\psi\yd_{-k}$, where $\tan(\theta_k) = k/\D$. 

We now express $\sigma^z$ in terms of Bogoliubov quasi-particles creation and annihilation: \be \sum_i \sigma^z_i = \sin(\theta_k) \gamma\yd_k \gamma\yd_{-k} + \cos(\theta_k) \gamma\yd_k \gamma_{k} + {\rm H. c.}.\ee 
The required matrix element is: \be M_{k,0} \equiv \bra{\psi_k} \sum_i\sigma^z \GS = \sin(\theta_k) \approx \frac{k}{\ve_k}.\ee Here $\ket{\psi_k}$ is a state with two Bogoliubov quasi-particles (at momentum $k$ and $-k$) and energy $2\ve_k \equiv 2\sqrt{k^2 + \D^2}$. The computation of the absorption spectrum is straightforward: 
\bea 
S_{\rm hop}(\w) &\approx& \sum_k \left| M_{k,0}\right|^2 \delta(\w-2\ve_k) \approx \left(\frac{k}{\ve_k}\right)^2 \left( \frac{\partial k}{\partial \ve_k} \right)_{\w=2\ve_k}\nonumber\\ 
&=&\left(\frac{k}{\ve_k}\right)_{\omega = 2\ve_k} = \frac{\sqrt{\omega^2-(2\D)^2}}{\omega}\Theta(\omega-2\D)\label{Sbragg}
\eea
Here $\Theta(\omega)$ is the Heaviside step function.

\bibliographystyle{aipnum4-1}
\bibliography{HIexcite}

\end{document}